\documentclass[prl,twocolumn,preprintnumbers,amsmath,amssymb]{revtex4}
\usepackage{graphicx}
\usepackage{graphics}
\usepackage{psfrag}
\usepackage{hyperref}

\begin{document}

\title{Three-body loss in lithium from functional renormalization}
\author{R. Schmidt}
\author{S. Floerchinger}
\author{C. Wetterich}
\affiliation{Institut f\"{u}r Theoretische Physik\\Universit\"at Heidelberg\\Philosophenweg 16, D-69120 Heidelberg, Germany}

\begin{abstract}
We use functional integral methods for an estimate of the three-body loss in a three-component $^6$Li ultracold atom gas. We advocate a simple picture where the loss proceeds by the formation of a three-atom bound state, the trion. In turn, the effective amplitude for the trion formation from three atoms is estimated from a simple effective boson exchange process. The energy gap of the trion and other key quantities for the loss coefficient are computed in a functional renormalization group framework.
\end{abstract}


\maketitle
Ultracold fermion gases with three components show new features as compared to the well studies systems with two components. For degenerate fermions with SU(3) symmetry one finds in the unitarity limit of infinite scattering length the interesting tower of Efimov states \cite{Efimov}. These are series of three-atom bound states with a geometrically decreasing gap parameter, reflecting the violation of scale symmetry by a limit cycle scaling behavior of the renormalization flow \cite{Bedaque,FSMW}. In the vicinity of the Fesh\-bach resonance the lowest three-atom state, the trion, is below the open channel energy level. At low temperature one can infer a phase structure different from the BCS-BEC crossover for the two-component system, namely an intermediate trion phase without superfluidity separating the superfluid BCS and BEC phases \cite{FSMW}. Interesting quantum phase transitions may describe the phase transition between phases at vanishing temperature.

\begin{figure}[h!]
\centering
\includegraphics[width=0.5\linewidth]{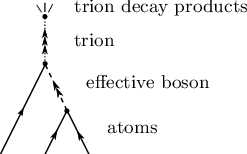}
\caption{Three-body loss process involving the trion.}
\label{fig:treeLossProcess}
\end{figure}

The trion bound state is also expected to persist if the SU(3) symmetry is violated by a different location and strength for the Feshbach resonances between different pairs of atomic components. Recent measurements of the three-body loss coefficient in a three-component system of $^6$Li \cite{Ottenstein, Huckans2008} may find an interpretation in this way \cite{Braaten:2008wd, Naidon}. We investigate here a simple setting, where the loss arises from the formation of an intermediate trion bound state, which subsequently decays into unspecified degrees of freedom -- possibly the ``molecule type'' dimers associated to the nearby Feshbach resonances. In turn, the trion formation from three atoms proceeds by the exchange of an effective bosonic field, as shown in Fig. \ref{fig:treeLossProcess}.
We estimate the loss coefficient $K_3$ as being proportional to
\begin{equation}
p=\left| \sum_{i=1}^3 \frac{h_i g_i}{m_{\phi i}^2}\frac{1}{\left(m_\chi^2-i \frac{\Gamma_\chi}{2}\right)}\right|^2.
\label{eq:decayprob}
\end{equation}
Here $m_\chi^2$ and $\Gamma_\chi$ are the trion gap parameter and decay width, while $m_{\phi i}^2$ describes a type of gap parameter for the effective boson, such that its propagator can be approximated by $m_{\phi i}^{-2}$. The Yukawa couplings $h_i$ couple the fermionic atoms to the effective boson, and the trion coupling $g_i$ accounts for the coupling between trion, atom and effective boson. We sum over the ``flavor'' indices $i=1,2,3$. We will estimate $m_\chi^2$, $m_{\phi i}^2$, $h_i$ and $g_i$ from the non-perturbative renormalization flow which arises from a simple truncation of the exact flow equation for the average action or flowing action \cite{Wetterich1993}, for reviews see \cite{Reviews}.

Recently we used the method of functional renormalization to describe a SU(3) invariant system of three fermion species close to a common Feshbach resonance \cite{FSMW, Moroz2008}. In this context we explored the manifestation of the Efimov effect and formulated some predictions on the quantum phase diagram in such systems. In contrast to this theoretical model, the system consisting of three-component $^{6}$Li atoms, which is of current experimental interest \cite{Ottenstein,Huckans2008}, does not possess this SU(3) symmetry. The main difference is that the resonances do not occur at the same magnetic field, and thus, for a given magnetic field $B$, the scattering lengths of different pairs of atoms, $(1,2)$, $(2,3)$, and $(3,1)$ differ from each other.

\begin{figure}[h!]
\centering
\includegraphics[width=0.75\linewidth]{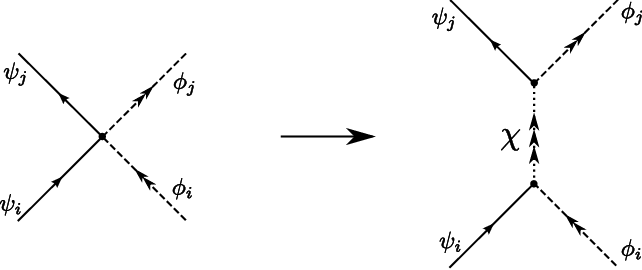}
\caption{Interaction between atoms $\psi$ and effective bosons $\phi$ as mediated by the trion field $\chi$.}
\label{fig:tree1}
\end{figure}

In this letter we generalize the model presented in \cite{FSMW} to cope with this more general situation. Our truncation of the (euclidean) average action then reads
\begin{eqnarray}
\nonumber
\Gamma_k &=& \int_x {\bigg \{} \psi_i^*(\partial_\tau-\Delta-\mu)\psi_i\\
\nonumber
&& +\phi_i^*\left[A_{\phi i}(\partial_\tau-\Delta/2)+m_{\phi i}^2\right]\phi_i\\
\nonumber
&& + \chi^*\left[\partial_\tau-\Delta/3+m_\chi^2\right]\chi\\
\nonumber
&& +h_i \epsilon_{ijk}(\phi_i^* \psi_j\psi_k-\phi_i\psi_j^*\psi_k^*)\\
&&+ g_i(\phi_i^* \psi_i^* \chi-\phi_i \psi_i \chi^*) {\bigg \}},
\label{eq:action}
\end{eqnarray}
where we choose natural units $\hbar=2M=1$, with the atom mass $M$. We sum over the indices $i$, $j$, $k$ wherever they appear. Here $\psi_i$ denotes the fermionic atoms, $\phi_i$ a bosonic auxiliary field which mediates the four-fermion interaction and $\chi$ is a fermionic field representing the bound state of three atoms. Formally, this trion field is introduced as the field mediating the interaction between atoms $\psi$ and bosons $\phi$. We show this schematically in Fig. \ref{fig:tree1}. In the limit $m_\chi^2\to\infty$, $m_{\phi i}^2\to\infty$, $h_i^2/m_{\phi i}^2\to|\lambda_i|$, $g_i^2/m_\chi^2\to |\lambda^{(3)}|$ the action describes pointlike two-body interactions with strength $\lambda_i$, as well as a three-body interaction with strength $\lambda^{(3)}$. We will concentrate on a microscopic interaction of this pointlike type. We consider the ``vacuum limit'' where temperature and atom density go to zero. Then the chemical potential $\mu$ in Eq. \eqref{eq:action} satisfies $\mu\leq 0$. A negative chemical potential $\mu$ has the meaning of an energy gap for the fermions when some other particle (boson or trion) has a lower energy. The dominant difference to the SU(3) symmetric model arises from the different propagators of the bosonic fields $\phi_1\widehat{=}\psi_2\psi_3$, $\phi_2\widehat{=}\psi_3\psi_1$, and $\phi_3\widehat{=}\psi_1\psi_2$. In addition, we allow in general for different Yukawa couplings $h_i$ corresponding to different widths of the three resonances. Also the Yukawa-like coupling $g_i$ that couples the different combinations of fermions $\psi_i$ and bosons $\phi_i$ to the trion field $\chi\widehat{=}\psi_1\psi_2\psi_3$ is permitted to vary with the species involved. Although the SU(3) symmetry is explicitly broken, the system exhibits three global U(1) symmetries corresponding to the three conserved numbers of species of atoms.

The renormalization flow of the various couplings from the microscopic (UV), $k=\Lambda$, to the physical, macroscopic (IR) scale, $k=0$, is obtained by inserting the ``truncation'' \eqref{eq:action} into the exact flow equation \cite{Wetterich1993}, for details we refer again to \cite{FSMW}. The flow equations for the two-body sector, i.~e. for the boson propagator parameterized by $A_{\phi i}$ and $m_{\phi_i}^2$, are very similar to the SU(3) symmetric case ($t=\ln(k/\Lambda)$)
\begin{eqnarray}
\nonumber
\partial_t A_{\phi i} &=& \frac{h_i^2 k^5}{6\pi^2 (k^2-\mu)^2},\\
\partial_t m_{\phi i}^2 &=& \frac{h_i^2 k^5}{6\pi^2 (k^2-\mu)^3}.
\label{eq:flowbosonprop}
\end{eqnarray}
Since the Yukawa couplings $h_i$ are not renormalized,
\begin{equation}
\partial_t h_i=0,
\label{eq:flowofh}
\end{equation}
we can immediately integrate the equations \eqref{eq:flowbosonprop}. The solution can be found in \cite{FSMW}. The microscopic values $m_{\phi i}^2(\Lambda)$ (bare couplings) have to be choosen such that the physical scattering lengths (at $k=0$) between two fermions (renormalized couplings) are reproduced correctly. They are given by the exchange of the boson field $\phi$. For example, the scattering length between the fermions 1 and 2 obeys
\begin{equation}
a_{12}=-\frac{h_3^2}{8\pi m_{\phi 3}^2},
\label{mphifixing}
\end{equation} 
where all ``flowing parameters'' are evaluated at the macroscopic scale $k=0$ and for $\mu=0$. We use this description for the scattering between fermions $\psi$ in terms of a composite boson field $\phi$ also away from the resonance. We emphasize that the field $\phi$ is not related to the closed channel Feshbach molecules of the nearby resonance. It rather describes an additional ``effective boson'' which may be seen as an auxiliary or Hubbard-Stratonovich field, allowing for a simple but effective description. For the numeric calculations in this note we will use large values of $h_i^2$ on the initial scale $\Lambda$. This corresponds to pointlike atom-atom interactions in the microscopic regime. 

Quite similar to the scattering between fermions $\psi$ in terms of the bosonic composite state $\phi$ we use a description of the scattering between fermions $\psi$ and bosons $\phi$ in terms of the trion field $\chi$. As an example, a process where the fermion $\psi_1$ and the boson $\phi_1$ scatter to a fermion $\psi_2$ and a boson $\phi_2$, is given by a tree level diagram as in Fig. \ref{fig:tree1}. For vanishing center-of-mass momentum the effective atom-boson coupling reads
\begin{equation}
\lambda_{1,2}^{(3)}=-\frac{g_1 g_2}{m_\chi^2}.
\end{equation}
The flow equations for the three-body sector within our approximation are given by the flow of the ``mass term'' for the trion field
\begin{equation}
\partial_t m_\chi^2 = \sum_{i=1}^3 \frac{2 g_i^2 k^5}{\pi^2 A_{\phi i}(3k^2-2\mu+2m_{\phi i}^2/A_{\phi i})^2}
\label{eq:flowofm}
\end{equation}
and the Yukawa-like coupling $g_i$ with flow equation
\begin{eqnarray}
\nonumber
\partial_t g_1  &=& -\frac{g_2 h_2 h_1 k^5\left(6k^2-5\mu+\frac{2m_{\phi2}^2}{A_{\phi 2}}\right)}{3\pi^2 A_{\phi 2}(k^2-\mu)^2\left(3k^2-2\mu+\frac{2m_{\phi2}^2}{A_{\phi2}}\right)^2}\\
&&-\frac{g_3 h_3 h_1 k^5\left(6k^2-5\mu+\frac{2m_{\phi3}^2}{A_{\phi 3}}\right)}{3\pi^2 A_{\phi 3}(k^2-\mu)^2\left(3k^2-2\mu+\frac{2m_{\phi3}^2}{A_{\phi3}}\right)^2}.
\label{eq:flowofg}
\end{eqnarray}
The flow equations for $g_2$ and $g_3$ can be obtained from Eq. \eqref{eq:flowofg} by permuting the indices $1$, $2$, $3$. 
For simplicity, we neglected in the flow equations \eqref{eq:flowofm} and \eqref{eq:flowofg} a contribution that arises from box-diagrams contributing to the atom-boson interaction. As described in \cite{FSMW} this term can be incorporated into our formalism using scale-dependent fields. Also terms of the form $\psi_i^*\psi_i \phi^*_j\phi_j$ with $i\neq j$, that are in principle allowed by the symmetries are neglected by our approximation in Eq. \eqref{eq:action}. We expect that their quantitative influence is sub dominant as it is the case for the SU(3) symmetric case \cite{Moroz2008}.

\begin{figure}[h]
\centering
\includegraphics[width=\linewidth]{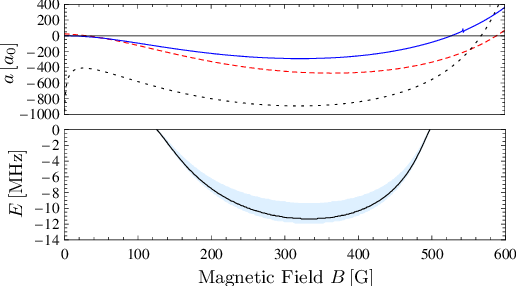}
\caption{(Color online) {\itshape Upper panel:} Scattering length $a_{12}$ (solid), $a_{23}$ (dashed) and $a_{31}$ (dotted) as a function of the magnetic field $B$ for $^6$Li. These curves were calculated by P.~S.~Julienne \cite{Bartenstein} and taken from Ref. \cite{Ottenstein}.\\
{\itshape Lower panel:} Binding energy per atom $E$ of the three-body bound state $\chi\widehat{=}\psi_1\psi_2\psi_3$. The solid line corresponds to the initial value $h^2=100 a_0^{-1}$, while the shaded region gives the result in the range $h^2=20 a_0^{-1}$ (upper border) to $h^2=300 a_0^{-1}$ (lower border).}
\label{fig:Energies}
\end{figure}

We apply our formalism to $^6$Li by choosing the initial values of $m_{\phi i}^2$ at the scale $\Lambda$ such that the experimentally measured scattering lengths (see Fig. \ref{fig:Energies}) are reproduced. For $A_{\phi i}(\Lambda)=1$, the value of $h_i$ parameterizes the momentum dependence of the interaction between atoms on the microscopic scale.  Close to the Feshbach resonance it is also connected to the width of the resonance $h_i^2\sim\Delta B$.  We choose here equal and large values for all three species $h_1=h_2=h_3=h$. This correspond to pointlike interactions at the microscopic scale $\Lambda$. Since the precise value of $h$ is not known, we use the dependence of our results on $h$ as an estimate of their uncertainty. 
The initial values of the couplings $m_\chi^2$ and $g_i$ are parameters in addition to the scattering lengths which have to be fixed from experimental observation. For equal interaction between atoms $\psi$ and bosons $\phi$ in the UV, the parameter to be fixed is
\begin{equation}
\lambda^{(3)}=-\frac{g^2(\Lambda)}{m_\chi^2(\Lambda)}
\end{equation}
with $g=g_1=g_2=g_3$. Pointlike interactions at the microscopic scale may be realized by $m_\chi^2(\Lambda)\to \infty$. 

We solve the flow equations \eqref{eq:flowbosonprop}, \eqref{eq:flowofh}, \eqref{eq:flowofm} and \eqref{eq:flowofg} numerically. For some range of $\lambda^{(3)}$ and $\mu\leq 0$ we find $m_\chi^2=0$ at $k=0$ for large enough values of the scattering lengths $a_{12}$, $a_{23}$ and $a_{31}$. This indicates the presence of a bound state of three atoms $\chi\widehat{=}\psi_1\psi_2\psi_3$. The binding energy per atom $E$ of this bound state is given by the chemical potential $|\mu|$ with $\mu$ fixed such that $m_\chi^2=0$ \cite{FSMW}. To compare with the recently performed experimental investigations of $^6$Li \cite{Ottenstein, Huckans2008}, we adapt the initial value $\lambda^{(3)}$ such that the appearance of this bound state corresponds to a magnetic field $B=125 \text{G}$, the point where strong three-body losses have been observed. Using the same initial value of $\lambda^{(3)}$ also for other values of the magnetic field, all microcsopic parameters are now fixed. We can now proceed to the predictions of our model.

First we find that the bound state of three atoms exists in the magnetic field region from $B=125 \text{G}$ to $B=498  \text{G}$. The binding energy per atom $E$ is plotted in the lower panel of Fig. \ref{fig:Energies}. We choose here $h^2=100\, a_0^{-1}$, as appropriate for $^6$Li in the (1,2)-channel close to the resonance, while the shaded region corresponds to $h^2\in(20\, a_0^{-1},300\, a_0^{-1})$.

As a second prediction, we present an estimate of the three-body loss coefficient $K_3$ that has been measured in the experiments by Jochim {\itshape et al.} \cite{Ottenstein} and O'Hara {\itshape et al.} \cite{Huckans2008}. For this purpose it is important to note that the fermionic bound state particle $\chi$ might decay into states with lower energies. These may be some deeply bound molecules not included in our calculation here. We first assume that such a loss process does not depend strongly on the magnetic field $B$ and therefore work with a constant decay width $\Gamma_\chi$ for the bound state $\chi$. 
The decay width $\Gamma_\chi$ appears as an imaginary part of the trion propagator when continued to real time
\begin{equation}
G_\chi^{-1}=\omega-\frac{\vec p^2}{3}-m_\chi^2+i\frac{\Gamma_\chi}{2}.
\end{equation}
Instead of working with negative $\mu$ chosen such that $m_\chi^2=0$, as done for the computation of the binding energy, we now perform an energy shift such that the zero energy level corresponds to the open channel and therefore $\mu=0$. In the region from $B=125 \text{G}$ to $B=498 \text{G}$ the energy gap of the trion is then negative $m_\chi^2<0$.

The three-body loss coefficient $K_3$ for arbitrary $\Gamma_\chi$ is obtained as follows. The amplitude to form a trion out of three fermions with vanishing momentum and energy is given by $\sum_{i=1}^3 h_i g_i/m_{\phi i}^2$.
The amplitude for the transition from an initial state of three atoms to a final state of the trion decay products (cf. Fig \ref{fig:treeLossProcess}) further involves the trion propagator that we evaluate in the limit of small momentum $\vec p^2=(\sum_i \vec p_i)^2\to0$, and small on-shell atom energies $\omega_i=\vec p_i^2$, $\omega=\sum_i \omega_i\to 0$. A thermal distribution of the initial momenta will induce some corrections. Finally, the loss coefficient involves the unknown vertices and phase space factors of the trion decay -- for this reason our computation contains an unknown multiplicative factor $c_K$. In terms of $p$ given by Eq. \eqref{eq:decayprob} we obtain the three-body loss coefficient
\begin{equation}
K_3=c_K \, p.
\end{equation}

\begin{figure}[h!]
\centering
\includegraphics[width=\linewidth]{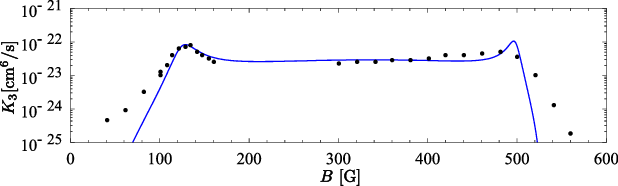}
\caption{(Color online) Loss coefficient $K_3$ in dependence on the magnetic field $B$ as measured in \cite{Ottenstein} (dots). The solid line is a two-parameter fit of our model to the experimental curve. We use here a decay width $\Gamma_\chi$ that is independent of the magnetic field $B$.}
\label{fig:Losscoefficient}
\end{figure}

Our result as well as the experimental data points \cite{Ottenstein} are shown in Fig. \ref{fig:Losscoefficient}. The agreement between the form of the two curves is already quite remarkable.

We have used three parameters, the location of the resonance at $B_0=125\text{G}$, the overall amplitude $c_K$ and the decay width $\Gamma_\chi$. They are essentially fixed by the peak at $B_0=125\text{G}$. The extension of the loss rate away from the peak involves then no further parameter.

Our simple prediction involves a rather narrow second peak around $B_1\approx500\text{G}$, where the trion energy becomes again degenerate with the open channel, cf. Fig. \ref{fig:Energies}. The width of this peak is fixed so far by the assumption that the decay width $\Gamma_\chi$ is independent of the magnetic field. This may be questionable in view of the close-by Feshbach resonance and the fact that the trion may actually decay into the associated molecule-like bound states which have lower energy. We have tested several reasonable approximations, which indeed lead to a broadening or even disappearance of the second peak, without much effect on the intermediate range of fields $150\text{G}<B<400\text{G}$.

In conclusion, a rather simple trion exchange picture describes rather well the observed enhancement of the three-body loss coefficient in a range of magnetic fields between $100\text{G}$ and $520\text{G}$. A similar trion dominated three-body loss is possible for large $B$ ($B\gtrsim850\text{G}$), where also a trion bound state with energy below the open channel exists. However, the dimer bound states are now above the open channel level, such that the trion decay may be strongly altered. The role of trion bound states in the resonance region is an interesting subject by its own, that can be explored by our functional renormalization group methods with an extended truncation.

\paragraph{Acknowledgement}
We thank S.~Jochim and the members of his group for interesting discussions and sending us their experimental data. We also thank S.~Moroz and J.~M.~Pawlowski for stimulating discussions and collaboration.

\end{document}